\documentstyle{l-aa}    % formato AA
\def\ltsim{\raise 2pt \hbox {$<$} \kern-1.1em \lower 4pt \hbox {$\sim$}}
\def\ltapprox{\raise 2pt \hbox {$<$} \kern-1.1em \lower 5pt \hbox {$\approx$}}
\def\gtsim{\raise 2pt \hbox {$>$} \kern-1.1em \lower 4pt \hbox {$\sim$}}
\def\gtapprox{\raise 2pt \hbox {$>$} \kern-1.1em \lower 5pt \hbox {$\approx$}}

\def\skuno{\vskip 20pt}

\def\p0{\phantom{0}}
\def\phns{\phantom{N-S }}
\def\phew{\phantom{E-W }}
\def\ph1{\phantom{1}}
\begin{document}
\thesaurus{03(11.09.1 IC310; 11.09.1 NGC1265; 11.09.1 3C129; 11.09.1 3C465; 
11.13.2; 13.18.1)}
\title
{Electron Ageing and Polarization in Tailed Radio Galaxies}
\skuno
\skuno
\author{L. Feretti\inst{1}  \and G. Giovannini\inst{1,2} \and U. Klein\inst{3} 
\and K.-H. Mack\inst{1,3} \and L.G. Sijbring\inst{4} \and G. Zech\inst{3}}
\offprints{lferetti@ira.bo.cnr.it}
\institute{
Istituto di Radioastronomia del CNR, Via P. Gobetti 101, I-40129 Bologna, 
    Italy
\and
Dipartimento di Astronomia, Universit{\`a} di Bologna, Via Zamboni 33, I-40126
    Bologna, Italy
\and
Radioastronomisches Institut der Universit\"at Bonn, Auf dem H\"ugel 71,
    D-53121 Bonn, Germany
\and
Kapteyn Astronomical Institute, Postbus 800, NL-9700 AV Groningen,
    The Netherlands}

\maketitle

\begin{abstract}

High-frequency observations of the tailed radio galaxies IC\ts310, NGC\ts1265, 
3C\ts129, and 3C\ts465 have 
been performed with the Effelsberg 100-m telescope. 
For the radio galaxies IC\ts310, NGC\ts1265 and 3C\ts465,  
radio data obtained at low frequencies with the Westerbork Synthesis 
Radio Telescope are also available. These new radio data allow us to map the 
extended structure of the radio galaxies and obtain spectral and polarization 
information in the outermost regions. 

The multi-frequency spectra were used to study the synchrotron ageing of
relativistic electrons with increasing distance from the active nucleus. We 
found that the spectrum in each radio galaxy progressively steepens with 
distance, and at each location it is steeper at high frequencies. The spectra 
are fitted by models involving synchrotron energy losses and 
the critical frequency is obtained at increasing distance from 
the core. Assuming that the magnetic field is the equipartition value, we 
obtain the radiating electron lifetimes and consequently their drift 
velocities. Our results imply the existence of reacceleration processes 
or bulk motions along the tails. 

The polarization data at 10.6~GHz give information on the intrinsic degree of
polarized flux and the orientation of the magnetic field. We find that the
polarization percentage increases along the tails, reflecting an increase of
the degree of ordering of the magnetic field. The magnetic field in the tails
is longitudinal. 


\end{abstract}

\section{Introduction}

The study of extended radio galaxies in clusters of galaxies is important to
understand the evolution of radio sources. In particular, the low-brightness
lobes are at a late stage of evolution and are typically confined by the
intergalactic medium, therefore synchrotron losses dominate in these regions.
Studies of extended radio sources often suffer from limitations of the
high-resolution interferometric observations, which can properly map the
low-brightness regions only at low frequencies (610 -- 327~MHz or lower). 
On the
other hand, information on the relativistic particle ageing and on the
polarization behaviour of the extended low-brightness regions is only obtained
with multiple frequencies. 

The Effelsberg 100-m radio telescope is at present the best available
instrument to map the outermost regions of extended radio galaxies with good
angular resolution, overcoming the  problem of the lack of short spacings 
(see e.g. Andernach et al. 1992 and Mack et al. 1993). Thanks to its high 
sensitivity, it is possible to map the extended structures at 10.6~GHz. At 
this high frequency, Faraday effects are generally negligible, thus the 
intrinsic degree of polarization and the orientation of magnetic field 
can be derived. 
Moreover, energy losses in relativistic electrons are severely affecting 
the emission spectrum, allowing a determination of the particle ageing at 
different distances from the core. 

In this paper we present new multi-frequency radio data of the extended 
tailed radio galaxies IC\ts310, NGC\ts1265, 3C\ts129, and 3C\ts465, revealing
the spectral index trends and the polarization behaviour 
up to the outermost regions. The spectral steepening can thence be used to 
determine the particle ages, following e.g. the procedure of Carilli et al. 
(1991). Tailed radio sources are particularly suited for this purpose. In 
fact, according to currently accepted models, the morphology of these sources 
is affected  by the drag action exerted by the intergalactic medium on the 
moving galaxy, and the low-brightness tails consist of aged relativistic 
electrons, which have reached pressure equilibrium with the ambient gas and 
are loosing energy essentially through synchrotron radiation and the inverse 
Compton effect. Since the radio emitting electrons are expected to be older 
at greater distance from the nucleus, the variation of the physical conditions 
with distance is related to the time evolution of the radio plasma. 

\begin{table*}
\caption{Radio sources and observational details}
\begin{flushleft}
\begin{tabular}{lcclrccccc}
\hline
\noalign{\smallskip}
Source &  RA(B1950) &  DEC(B1950) & \ \ z & Freq. &
\# Cov.  &  Map size & \multicolumn{1}{c}{HPBW} & $\sigma_{\rm I}$ 
& $\sigma_{\rm U,Q}$  \\
& [\ \ $^{\rm h}$ \ \ $^{\rm m}$ \ \ $^{\rm s}$] 
& [\ \ $^{\circ}$ \ \ $\arcmin \ \ \arcsec$]& &[MHz]
& & &\multicolumn{1}{c}{[$\arcsec$]}& [mJy/beam] & [mJy/beam] \\
\noalign{\smallskip}
\hline
\noalign{\smallskip}
IC\ts310 & 03 \ 13 \ 25.2 & 41 \ 08 \ 30 & 0.0183  
&  327  & - & -   &      69   &  3.2  &  -   \\
      &  &  &         &  610   & - & -    &  69   &  0.4  &  -   \\
      &  &  &         &  10550  & 12 &  $29\arcmin \times 8\arcmin$ 
&  69   &  0.5  &  0.3 \\
NGC\ts1265 & 03 \ 14 \ 56.8 & 41 \ 40 \ 32 & 0.0183  
&  327  & - &  -  &      147  &  5.1  &  -   \\
        & & &         &  610   & - & -      &  147  &  0.3  &  -   \\
%        & & &         &  2695  & &       &  258  &  5.2  &  3.5 \\
        & & &         &  4750 &  13 &  $28\arcmin \times 20\arcmin$ 
&  147  &  5.0  &  2.5 \\
        & & &         &  10550  & 22 & $35\arcmin \times 14\arcmin$ 
&  69   &  0.6  &  0.5 \\
3C\ts129   & 04 \ 45 \ 21.0 & 44 \ 56 \ 48 & 0.021   
&  2695  & 30 & $40\arcmin \times 26\arcmin$   &      258  &  4.4  &  3.3 \\
        & & &         &  4750  &  30 & $35\arcmin \times 17\arcmin$    
&  147  &  3.4  &  1.5 \\
        & & &         &  10550  & 21 & $46\arcmin \times 25\arcmin$      
&  69   &  1.0  &  0.4 \\
3C\ts465   & 23 \ 35 \ 59.0 & 26 \ 45 \ 16 & 0.0322  
%&  2695  & - &  -  &  258  &  5.0  &  2.4 \\
     & 610  & - & - & 147 &   5.0 & - \\
        & & &         &   4750  & 11 & $26\arcmin \times 18\arcmin$ 
&  147  &  2.3  &  1.3 \\
        & & &         &  10550  & 20 & $35\arcmin \times 14\arcmin$ 
&  69   &  0.9  &  0.5 \\
\noalign{\smallskip}
\hline
\noalign{\smallskip}
\end{tabular}
\end{flushleft}
\end{table*}

In Sect.~2 we briefly describe the observations and data reduction techniques. 
Section~3 
presents the results, which are discussed in Sect.~4, with emphasis on
the particle lifetime and velocity, and the magnetic field morphologies. In
Sect.~5  we summarize our results and present our conclusions. A Hubble 
constant H$_0$ = 100~km~s$^{-1}$~Mpc$^{-1}$ and a q$_0= 1$ are assumed.

\section{Observations and data analysis}

The observations at 10.6, 4.8, and 2.7~GHz have been carried out 
using the 100-m radio telescope at  Effelsberg. 
In Table~1 we have listed the radio galaxies, together with their position
and redshift, the observing frequency, the angular resolution
in the final maps, the r.m.s. noise level in the I and U, Q maps.

Fields were mapped in azimuth and elevation (see Emerson et al. 1979
for details) for each source ensuring sufficiently large areas to also 
account for the beam throws envolved. The number of coverages and field sizes
for each source are given in Table.~1.
The 2.7-GHz map of 3C\ts129  was obtained by scanning in right ascension
and declination 30 times, as this receiver has a single feed so that 
mapping in Az-El is not mandatory. At the two highest frequencies,
difference maps were obtained to remove atmospheric fluctuations in
the signals and the conventional restoration technique of Emerson et al. 
(1979) was applied. The resulting Stokes I, U, and Q maps were subsequently
transformed into the equatorial coordinate system, and individual coverages
finally averaged (with weights proportional to the inverse squares of the 
rms noise values). Maps at 10.6 GHz were finally CLEANed, 
with the algorithm described by  Klein \& Mack (1995). 
Calibration of the flux density scale and the polarization 
parameters was achieved by frequently cross-scanning
and mapping the point sources 3C~48, 3C~138, 3C~286, and 3C~295, 
with the flux density scale adopted from Baars et al. (1977).

For the radio galaxies IC\ts310 and NGC\ts1265, maps at 610 and 327~MHz were 
extracted from the image of the Perseus cluster (Sijbring 1993), obtained 
with the Westerbork Synthesis Radio Telescope (WSRT). The shortest baseline at 
both frequencies was 36~m, therefore only structures larger than 48$^{\prime}$ 
at 610~MHz and $1\fdg4$ at 327~MHz are missed. This ensures that these 
measurements are not affected by the lack of short baselines, and a comparison 
between Effelsberg and WSRT data is possible. 
Similarly, the WSRT map of the radio galaxy 3C\ts465, obtained
by J\"agers (1987) with the WSRT at 610 MHz, was used in the present work.  
The WSRT maps, available only in 
total intensity, were convolved to the resolution of the Effelsberg maps. 

The images were analyzed with the AIPS package. 
To compare observations of the same source at different frequencies
we convolved all the maps to the lowest resolution with a gaussian.
Thereafter, the maps were re-gridded by means of the task `HGEOM' in AIPS to 
have identical pixel size.

\section{Results}

\subsection{IC\ts310 - 0313+411}

IC\ts310 is located in the Perseus cluster. Previous observations of this radio
galaxy are listed by Mack et al. (1993), who first mapped the large-scale
structure of this source at a high frequency. The 10.6-GHz map presented here 
in Fig.~1 has a higher sensitivity, owing to additional scans. For the spectral
comparison, we used the WSRT maps at 327 and 610~MHz by Sijbring (1993),
convolved to the same resolution of 69$^{\prime\prime}$. 

The trend of the spectral index ($\alpha^{0.6}_{0.3}$ and
$\alpha^{10.6}_{0.6}$; I$_{\nu}\propto\nu^{-\alpha}$) derived along the maximum brightness ridge is given in
Fig.~5. The low-frequency spectral index $\alpha^{0.6}_{0.3}$ is 0.4, almost
constant up to 2$^{\prime}$ from the core, suggesting that the particles 
radiating in this region and at low frequencies have not suffered synchrotron 
losses. A clear steepening is visible at high frequencies and with increasing 
distance from the core. 

The tail is strongly polarized, up to a distance of 3$^{\prime}$ from the core,
with the polarization percentage increasing along the tail (see Fig.~6). The
orientation of the polarization vectors is perpendicular to the tail (see Fig.~1). 

\subsection{NGC\ts1265 - 0314+416}

NGC\ts1265 is located in a peripheral region of the Perseus Cluster, at about
27$^{\prime}$ north-west of the active galaxy NGC\ts1275 (3C\ts84). 
Previous studies
of this source are listed in Mack et al. (1993), who obtained the first image
at 10.6~GHz. The 10.6-GHz map presented here (Fig.~2) has a much higher
sensitivity than that of Mack et al. (1993), owing to a longer observing time.
We also obtained observations at 
4750~MHz and extracted WSRT maps at 327 and 610~MHz from the
Perseus Cluster image obtained by Sijbring (1993). 
We note that the very-low-brightness feature first reported by Gisler \& Miley (1979) and observed 
with higher sensitivity and dynamic range by Burns et al. (1992) and by 
Sijbring (1993) is just indicated in our 10.6-GHz map if smoothed
to a 1\farcm5 beam, especially in the (more sensitive) map of linear 
polarization. At the other frequencies the feature, if present, is 
concealed by the sidelobe structure which we could not yet clean at this
wavelength. 

All maps 
were smoothed to the common resolution of 147$^{\prime\prime}$ for a
multifrequency comparison.
The trend of the spectral index along the ridge of maximum brightness for the 3
frequency pairs is presented in Fig.~5. Between 0.3 and 4.8~GHz it is in the
range 0.5 -- 0.8, while beyond 4.8~GHz a strong steepening is evident, which
increases with distance from the core, reaching a value $\alpha >$ 2 at
the tail's end. 

Strong linear polarization is found in the external tail region (see Fig.~6).
The polarization percentage at 10.6~GHz increases along the tail up to
$\sim$40\%. At 4.8~GHz, the degree of polarization is comparable to that at
10.6~GHz up to 5$^{\prime}$ from the core, and is lower beyond that distance. 
The electric vector at both frequencies is perpendicular to the tail 
orientation. 
%At 2.7~GHz the polarization data are inconclusive, because of 
%the large observing beam.

\subsection{3C\ts129}

This radio source is identified with an E galaxy in the galactic plane ({\it
l}$_{II}$=160.4$^{\circ}$, {\it b}$_{II}$=0.1$^{\circ}$). It was studied by
many authors (e.g. Miley 1973, Rudnick \& Burns 1981, Van Breugel 1982, Van
Breugel \& J\"agers 1982) with interferometric observations. The Effelsberg 
map at 10.6~GHz is given in Fig.~3. Maps were also obtained at 2.7 GHz
and 4.8 GHz. The tail of this source, 
which at lower frequencies is about 30$^{\prime}$ long (Van Breugel 1982), is 
detected out to about 20$^{\prime}$ at 2.7 GHz and to about
12$^{\prime}$ at 10.6~GHz. 

The spectral index along the source was obtained with an angular resolution
of about 4$^{\prime}$ (Fig.~5). The spectrum strongly steepens from the core to
the outer regions, reaching values up to $\alpha$$\sim$3. The polarization
percentage (Fig.~6) in the outermost tail region reaches values of
$\sim$60-70\% at high frequencies and is slightly lower at 2.7~GHz. The
polarization vector at 10.6~GHz is perpendicular to the tail, and shows
significant rotation at lower frequencies. 

\subsection{3C\ts465}

This source is identified with a cD galaxy at the centre of the rich cluster
A\ts2634. It is the prototype of wide-angle tailed radio sources 
(see e.g. Eilek et
al. 1984 and references therein), and a model of its dynamical evolution
has been proposed by Leahy (1984). High-resolution VLA maps show an unresolved
core, and two asymmetric jets which terminate in a bright spot and after 
a sharp bending, in extended lobes. In our maps at 4.8~GHz (not shown
here) and 10.6~GHz (Fig.~4), both the N-S and and E-W tails are well
resolved. We derived the point-to-point spectrum in the tails, 
with an angular resolution of 147$^{\prime \prime}$, using
also the 610 MHz WSRT map of J\"agers (1987).

The spectral index between 0.6 and 4.8 GHz is lower than that between
4.8 and 10.6 GHz, except for the outermost points of the N-S
lobe, where however the two values are  consistent within the
errors. A remarkable high-frequency steepening is present in the
E-W lobe, at increasing distance from the core (Fig.~5). 
The polarization percentage at both 10.6 and 4.8~GHz 
is asymmetric and is higher in the N-S lobe (Fig.~6). 
%At 2.7~GHz the 
%polarization data are inconclusive, owing the large observing beam. 

\section{Discussion} 

\subsection{Evolution of the spectrum along the sources}

The spectral behaviour in all the sources presented here is similar. The 
spectrum between two given frequencies shows a progressive steepening with 
increasing distance from the core. Moreover, the steepening is stronger in 
the high-frequency range than at lower frequencies. Since the radio emitting 
electrons are expected to be older at larger distance from the core, the 
variation of the spectrum with distance is related to its evolution with 
time under the effect of radiation losses.

\begin{table*}
\caption{Critical frequencies, electron lifetimes and projected
velocities for IC\ts310 with $\gamma$ = 1.8}
\begin{flushleft}
\begin{tabular}{crrrrrrr}
\hline
\noalign{\smallskip}
\multicolumn{1}{c}{Distance}& $\nu_{\rm c}$(KP) & $\nu_{\rm c}$(JP)& H$_{\rm eq}$ &Age(KP)&Age(JP)&Vel(KP)&Vel(JP) \\
$[\arcmin]$&[GHz] & [GHz] & [$\mu$G] & [$\times$10$^7$yr] & [$\times$10$^7$yr]
& [100 km s$^{-1}$] & [100 km s$^{-1}$] \\ 
\noalign{\smallskip}
\hline
\noalign{\smallskip}
  0.5  &    $>$30.0  &  $>$30.0  & 4.3 &  $<$2.0 & $<$2.0 & $>$3.8 & $>$3.8 \\
  1.0  &    $>$30.0  &  $>$30.0  & 3.6 &  $<$2.2 & $<$2.2 & $>$6.9 & $>$6.9 \\
  1.5  &     30.0  &   23.9  &   3.3   &  2.4   &  2.6 & 9.6 & 8.6 \\
  2.0  &     12.1  &  10.3   &   3.3   &  3.7   &  4.0 & 5.6 & 5.5 \\
  2.5  &      5.5  &   5.6   &   3.1  &   5.7   & 5.6  & 3.9 & 4.8 \\
  3.0  &      3.6  &   4.5   &   3.1  &   6.9   & 6.2  & 5.9 & 12.0 \\
  3.5  &      2.2  &   1.8   &   2.9  &   9.2   & 10.3 & 3.4 & 1.9  \\
  4.0  &      1.7  &   1.4   &   2.8  &   10.5   & 11.7 & 5.8 & 5.4 \\
  4.5  &      1.2  &   1.0   &   2.7  &   12.7   & 13.9 & 3.4 & 3.4 \\
  5.0  &      0.7  &   0.6   &   2.3  &   17.5   & 18.2 & 1.6 & 1.8 \\
  5.5  &      0.4  &   0.5   &   2.2  &   21.8   & 21.1 & 1.8 & 2.6 \\
  6.0  &      0.3  &   0.4   &   1.9  &   25.7   & 22.7 & 1.9 & 4.7 \\
\noalign{\smallskip}
\hline
\noalign{\smallskip}
\end{tabular}
\end{flushleft}
\end{table*}

\begin{table*}
\caption{Critical frequencies, electron lifetimes and projected velocities
for NGC\ts1265 with $\gamma$ = 2.2}
\begin{flushleft}
\begin{tabular}{crrrrrrr}
\hline
\noalign{\smallskip}
\multicolumn{1}{c}{Distance}& $\nu_{\rm c}$(KP) & $\nu_{\rm c}$(JP)  & H$_{\rm eq}$ & Age(KP) &  Age(JP) &Vel(KP)&Vel(JP) \\
$[\arcmin]$&[GHz]&[GHz] & [$\mu$G] & [$\times$10$^7$yr] & [$\times$10$^7$yr]
& [100 km s$^{-1}$] & [100 km s$^{-1}$]  \\ 
\noalign{\smallskip}
\hline
\noalign{\smallskip}
 \ph11    &  $>$30.0 & $>$30.0 & 3.5 & $<$2.3 & $<$2.3 & $>$6.6 & $>$6.6 \\
 \ph12    &  $>$30.0 & 26.0 & 3.5 & $<$2.3 & 2.5 & $>$13.2 & 12.3 \\
 \ph13    &  17.6  &  14.5  & 3.2  & 3.1 & 3.4 & 14.6 & 15.4 \\
 \ph14    &  12.9  &  11.0  & 3.1  & 3.7 & 4.0 & 26.5 & 26.9 \\
 \ph15    &  10.9  &   9.5  & 3.1  & 4.0 & 4.3 & 45.7 & 50.3 \\
 \ph16    &   9.1 &    8.2  & 3.0  & 4.4 & 4.7 & 35.1 & 38.7 \\
 \ph17    &   7.8  &   7.2  & 2.9  & 4.9 & 5.1 & 35.9 & 39.7 \\
 \ph18    &   6.5  &   6.2  & 2.7  & 5.5 & 5.6 & 25.6 & 29.0 \\
 \ph19    &   5.5  &   5.3  & 2.6  & 6.0 & 6.1 & 28.5 & 29.6 \\
 10    &   4.8  &   4.6  & 2.3  & 6.6 & 6.7    & 26.5 & 25.1 \\
\noalign{\smallskip}
\hline
\noalign{\smallskip}
\end{tabular}
\end{flushleft}
\end{table*}

\begin{table*}
\caption{Critical frequencies,  electron lifetimes and projected velocities
for 3C\ts129 with $\gamma$ = 2.4}
\begin{flushleft}
\begin{tabular}{crrrrrrr}
\hline
\noalign{\smallskip}
\multicolumn{1}{c}{Distance}& $\nu_{\rm c}$(KP) & $\nu_{\rm c}$(JP)  & H$_{\rm eq}$ & Age(KP) & Age(JP) &Vel(KP)&Vel(JP) \\
$[\arcmin]$&[GHz]&[GHz]&[$\mu$G]&[$\times$10$^7$yr]&[$\times$10$^7$yr] 
& [100 km s$^{-1}$] & [100 km s$^{-1}$] \\ 
\noalign{\smallskip}
\hline
\noalign{\smallskip}
  \ph11   &  $>$30.0 & $>$30.0 & 4.1 &  $<$2.1 & $<$2.1 & $>$8.4 & $>$8.4 \\
  \ph12   &  $>$30.0 & $>$30.0 & 3.9 &  $<$2.1 & $<$2.1 & $>$16.2 & $>$16.2 \\
  \ph13   &  $>$30.0 & $>$30.0 & 4.2 &  $<$2.0 & $<$2.0 & $>$25.6 & $>$25.6 \\
  \ph14   &    26.0 &   21.0 &  3.6 &    2.4   & 2.7 &  28.8 & 25.9  \\
  \ph15   &    13.0 &   11.5 &  3.1 &    3.7   & 3.9 & 13.7  & 14.1 \\
  \ph16   &     7.6 &   7.7  &  3.0 &    4.8   & 4.8 & 14.7  & 19.1  \\
  \ph17   &     5.3 &   6.3  &  2.8 &    5.9   & 5.4 & 15.4  & 26.1 \\
  \ph18   &     4.4 &   5.7  &  2.7 &    6.6   & 5.8 & 26.9  & 52.2 \\
  \ph19   &     3.4 &   5.0  &  2.6 &    7.5   & 6.2 & 18.5  & 39.1 \\
 10   &     2.9 &   4.5  &  2.4 &    8.4   & 6.7   & 20.5   & 35.1 \\
 11   &     2.5 &   3.5  &  2.3 &    9.1   & 7.6   & 23.3  & 18.7 \\
 12   &     N.D. &   2.8  &  2.1 &     -     & 8.6 & -   & 17.8 \\
 13   &     N.D. &   2.3  &  1.9 &     -     & 9.5 & -   & 18.1 \\
 14   &     N.D. &   1.9  &  1.6 &     -     & 10.3 & -  & 22.7 \\
\noalign{\smallskip}
\hline
\noalign{\smallskip}
\end{tabular}
\end{flushleft}
\end{table*}

\begin{table*}
\caption{Critical frequencies, electron lifetimes and projected velocities
for 3C\ts465 with $\gamma$ = 2.2}
\begin{flushleft}
\begin{tabular}{crrrrrrr}
\hline
\noalign{\smallskip}
\multicolumn{1}{c}{Distance}  & $\nu_{\rm c}$(KP) & $\nu_{\rm c}$(JP)  & H$_{\rm eq}$ & Age(KP) &  Age(JP) &Vel(KP)&Vel(JP) \\
$[\arcmin]$ & [GHz]& [GHz] & [$\mu$G] & [$\times$10$^7$yr] & [$\times$10$^7$yr]
& [100 km s$^{-1}$] & [100 km s$^{-1}$] \\ 
\noalign{\smallskip}
\hline
\noalign{\smallskip}
N-S 1   &   $>$30.0 &  $>$30.0 & 4.5 & $<$1.9 & $<$1.9 & $>$13.7 & $>$13.7 \\
\phns2  &   $>$30.0 &  $>$30.0 & 4.7 & $<$1.8 & $<$1.8 & $>$28.4 & $>$28.4 \\
\phns3  &   $>$30.0 &  $>$30.0 & 5.2 & $<$1.7 & $<$1.7 & $>$46.4 & $>$46.4 \\
\phns4  &   26.0 &   20.9 &  6.3 &   1.5   &  1.7    & 69.3 & 62.2 \\
\phns5  &   N.D. &   N.D. &  5.2 &   -   &  -  & - & - \\
\phns6  &   N.D. &   N.D.  &  4.9 &  -   &  -  & - & - \\
\phns7  &    N.D. &  N.D.  &  3.8 &  -   &  -  & - & - \\
\phns8  &    N.D. &   N.D.  &  2.7 &  -   &  -  & - & - \\
E-W 1   &   $>$30.0 & $>$30.0 &  5.0 &  $<$1.7 &  $<$1.7 & $>$14.9 & $>$14.9 \\
\phew2  &   $>$30.0 & $>$30.0 &  6.5 &  $<$1.3 &  $<$1.3 & $>$38.5 & $>$38.5 \\
\phew3  &   27.9 &   22.6 &  7.7 &   1.2   &  1.3  & 67.3 & 60.5 \\
\phew4  &   13.0 &   11.3 &  6.2 &   2.1   &  2.3  & 26.1 & 25.3 \\
\phew5  &    5.5 &    6.5 &  4.7 &   4.2   &  3.9  & 12.3 & 16.1 \\
\phew6  &    3.5 &    4.3 &  3.9 &   6.1   &  5.5  & 14.1 & 16.3 \\
\phew7  &    3.5 &    3.8 &  3.4 &   6.6   &  6.3  & 53.8 & 31.9 \\
\phew8  &    4.6 &    4.4 &  3.2 &   5.9  &  6.0   & -    & - \\
\noalign{\smallskip}
\hline
\noalign{\smallskip}
\end{tabular}
\end{flushleft}
\end{table*}

The synchrotron loss mechanism for an ensemble of electrons with an initial
power-law energy distribution N(E)dE $\propto$ E$^{-\gamma}$dE produces a
curvature of the originally straight power-law emission spectrum I($\nu$)
$\propto \nu^{-\alpha}$, with $\alpha$ = ($\gamma$ --1)/2. The curvature
manifests itself at the critical frequency $\nu_{\rm c}$, related to the 
electron age. The shape of the expected synchrotron spectrum can be computed
analytically as a function of the critical frequency, the electron energy
distribution index $\gamma$ and the evolution of the electron pitch-angle
distribution with time (Pacholczyk 1970). The model of Kardashev-Pacholczyk
(KP) is obtained in the case that electrons maintain the same pitch angle
throughout their radiative lifetime (Kardashev 1962). The model of
Jaffe-Perola (JP) assumes that there is a redistribution of electron pitch
angles on short time scales in comparison with their radiative lifetimes, due
to their scattering on magnetic field irregularities (Jaffe \& Perola 1973). 

We have fitted the spectra of our sources obtained at different distances from
the core (see previous section) using both the KP and JP models, since we do
not have observational evidence in favour of one of them. 
We also used different
values of the exponent $\gamma$ of the energy distribution function of the
electrons, when no information about this parameter could be obtained from the
low-frequency spectrum. 
The fitting procedure was developed by Murgia \& Fanti (1996).

In Fig.~7 we present an example of spectral fits. In Tables~2 to 5 we give
the break frequencies obtained from the fitting procedure. The values of 
the critical frequencies of the KP model have been divided by 4/9 for 
convenience, for the computation of ages (see next subsection). 
Typical errors of
the break frequencies are of the order of 400~MHz. 

We note that in the above analysis, the energy losses due to expansion are 
not considered. Their effect would be to increase the effect of 
synchrotron losses, without modifying the shape of the spectrum. In the case
of constant expansion rates, the values of the break frequencies would be
shifted by the same amount and the differential behaviour 
of $\nu_c$ along the sources would be unchanged. 
However, since low brightness lobes in tailed radio galaxies are likely 
to be confined by the outer medium (see e.g. Feretti et al. 1992), 
expansion losses should be negligible.

The results of our spectral fits 
do not allow us to argue in favour of either the KP or JP model.
They are discussed for each source below. 

{\bf IC\ts310} - The low-frequency spectral index of this source up to   
2$^{\prime}$
from the core is $\alpha^{0.6}_{0.3}$$\sim$0.4. Since we do not expect any
effect of self-absorption or ageing in this region we can assume that this
spectrum reflects the original energy distribution of the radiating electrons.
This implies an index $\gamma$ = 1.8. The data are well fit with both the KP
and the JP models. The break frequencies are given in Table~2. 

{\bf NGC\ts1265} - We fitted the spectra of this source at different distances
from the core using flux densities at 4 frequencies (see Fig.~5). The 
low-frequency spectrum within a few arcmin from the nucleus has a spectral 
index in the range of $0.4-0.6$, therefore only a range for the parameter 
$\gamma$ can be inferred. We have used $\gamma$ = 1.8, 2.0, and
2.2. The value 
of 1.8 gives a bad fit with both models. Fits with $\gamma$ =
2.0 and $\gamma$ = 2.2 yield good results to both the KP and JP model, but
$\chi^2$ is lower for $\gamma$ = 2.2 (Table~3). The JP model gives a slightly
better fit up to 5$^{\prime}$ from the core, while the KP model seems more
suitable to describe the steepening in the most external region. 

{\bf 3C\ts129} - No information is available for this source about the 
low-frequency spectrum, therefore we fitted our data assuming 4 values of
$\gamma$: 1.8, 2.0, 2.2, 2.4. The best agreement with the data and the JP and 
KP models have been obtained using $\gamma$ = 2.4, which corresponds to an 
intrinsic spectral index of $\alpha = 0.7$. The values of the derived break frequencies 
are given in Table~4. The KP model gives a slightly better fit for the 
points within 7$^{\prime}$ from the core but it is unable to fit the very 
steep spectrum of the more external points where the JP model fit is 
significantly better. 

{\bf 3C\ts465} - In the innermost source region, 
the spectral index $\alpha^{4.8}_{0.6}$ is 0.6, therefore the electron energy
index $\gamma$ is likely to be $\leq$ 2.2. We fitted the spectra
using $\gamma$ = 2.0 and 2.2, and found that the last value gives
a better fit (see Table~5). 
In the innermost points, both the KP and JP models provide a
good fit to the spectral data, while at larger distances from the core, 
the KP model is increasingly better. In the outermost points of the
N-S tail, no satisfactory fit to the spectra was found, unless using
a different electron energy index. This implies that a simple model
of synchrotron emission cannot account for the spectral flattening at high
frequency.

\subsection{Electron lifetimes and velocities} 

Synchrotron radiation losses lead to a steepening of the emission spectrum at
high frequencies with time. From the value of the critical frequency 
$\nu_{\rm c}$ and of the magnetic field, it is possible to get information on 
the lifetime of radiating electrons suffering synchrotron and inverse Compton
losses against the 3K-background radiation. As the magnetic field strength, we
use the equipartition value estimated at different locations along each source.
The calculation of the equipartition magnetic field involves a number of
assumptions (Pacholczyk 1970). We assume the same volume 
for relativistic particles and
magnetic field (filling factor $\phi$ = 1), the same energy in electrons
and protons (k = 1), and upper and lower cutoff
frequencies of 10~MHz and 100~GHz, respectively. 
The derived values of H$_{\rm eq}$ for each
source are given in the corresponding table. 

The age due to synchrotron plus inverse Compton losses is then expressed by the
following relation (Alexander \& Leahy 1987): 
        $$\rm t = 1590 \cdot [\nu_{\rm c} \cdot (1 + \rm z)]^{-0.5}
               \cdot {{\rm H_{\rm eq}^{0.5}} \over {\rm H_{\rm eq}^2 + \rm 
H_{\rm m}^2}}~~ \rm Myr~~~~(1) $$
where $\nu_{\rm c}$ is the break frequency in GHz, H$_{\rm eq}$ is the 
equipartition magnetic field in $\mu$G, and H$_{\rm m} = 3.25 \cdot 
(1+z)^2$ is the equivalent field strength of the 3K radiation in $\mu$G. 
We note here that non-uniformities in the particle and field distributions
in the sources can complicate the interpretation of the spectral curvature
(Wiita \& Gopal-Krishna 1990, Eilek \& Arendt 1996). 

The previous equation is valid for the JP model, where a continuous
isotropization of the pitch angle is allowed with time. In the KP model, where
particles will maintain their initial pitch-angle over time, the expression for
the critical frequency differs from that valid in the JP model
by $<$sin$^2\theta>^2$, i.e.
4/9 for an isotropic pitch-angle distribution. Therefore, the right-hand side
of Eq.~(1) must be divided by 4/9 when using break frequencies obtained
in the KP model. To use Eq.~(1) also for
the KP model, the KP break frequencies in Tables~2 to 5 have been divided by
4/9. The derived ages at different distances from the core for each source are
given in Tables~2 to 5 for the two models and are plotted in Fig.~8. Typical 
errors in lifetimes are of the order of 5\%. 
From the age, we obtained the 
projected drift velocity of emitting particles at 
different distances from the core. These are also reported in Tables~2
to 5.

{\bf IC\ts310} seems to be old, since the electron lifetimes in the outer 
regions appear to be 2--2.5~$\times~10^8$~yrs. This is a bit surprising as
the source is smaller than the others studied here. This could suggest the 
existence of strong projection effects, but this is in contrast with
the low velocity (180 km s$^{-1}$) of IC\ts310 with respect 
to the Perseus cluster.
% if IC 310 is oriente at a small 
% angle to the line-of-sight, its length is  larger than that 
% measured on the plane of the sky. 
We note that the radio structure 
shows no evidence of any transverse expansion. This suggests that the source 
is well confined by the external medium and that electrons have not suffered 
significant expansion losses, and can thus survive for such a long time. 
The projected velocity of radiating electrons decreases along the tail from 
$\sim$900~km/s to $\sim$200~km/s. Projection effects would increase these 
values. Miley (1973) pointed out that the radio plasma of IC\ts310, being
$\sim$10 times less luminous than NGC\ts1265, may have been ejected with less 
kinetic energy. This may comply with the relatively low drift velocity 
inferred for its tail. 

{\bf NGC\ts1265} reaches an age of 6.7~$\times~10^7$~yr. 
The average electron velocity is $\sim$3000~km/s, and increases with 
distance, up to  7$^{\prime}$ from the core.
The radial velocity of this galaxy with respect to the Perseus cluster
is 2055 km s$^{-1}$. We note that O'Dea \& Owen (1987) argued, based on 
the magnetic field in the jets, that NGC\ts1265 should be within 45$^{\circ}$
of the plane of the sky. If this is the case the total galaxy velocity
would be consistent with the average value estimated from spectral aging.

{\bf 3C\ts129} is $\sim10^8$ yrs old in the more external regions. We remember
that for the outer regions of this source 
only the JP model is able to fit the spectral index data.
The particle velocity is around 2000~km/s, showing variations along the
tail. It seems that there is an increase of velocity at about 
6$^{\prime}$--7$^{\prime}$ from the core, where a large bend in the radio 
structure occurs.
 
In {\bf 3C\ts465} the  particle lifetimes in the E-W lobe   reach 
6--6.5~$\times~10^7$~yrs and the average velocity of emitting 
electrons is $\sim$3500~km/s. The velocity decreases at 4$^{\prime}$--
6$^{\prime}$ from the core, then it increases in the outermost source
regions. In the N-S tail, the points where the break frequencies
have been estimated are too few to derive the trend of electron age and 
velocity. The velocity implied at 4$^{\prime}$ from the nucleus is about
6500 km s$^{-1}$. 

The ages obtained for the present galaxies are of the same order as 
in other objects for which similar studies were made, such as 3C\ts31,
3C\ts449 (Andernach et al. 1992), and NGC\ts4869 (Feretti et al. 1990). 
%The relativistic electron velocities in the outer lobes are higher than 
%expected, as we would expect speeds of some hundred km~s$^{-1}$, 
%either considering that they diffuse at the typical sound speed 
%or that their motion reflects the proper motion  of the parent galaxy 
% within the cluster.  
%
The relativistic electron velocities are higher than expected,
as we would expect that the particles diffuse at the typical sound
speed (some hundred km~s$^{-1}$) or that their motion reflects 
the proper motion  of the parent galaxy  within the cluster, as 
in  NGC\ts1265.
We  note that there could be a selection effect, in that tailed
radio sources with very long tails may be those associated with
cluster galaxies with high proper motion.
However, in all cases we detect significant local variations, 
with values up to 5000--6000 km s$^{-1}$, at large distance from the core.
%We find, instead,  velocities up to a few thousands 
%km~s$^{-1}$, often increasing with distance. 
The inferred trends could be explained by:  a) projection effects, b) the
existence of reacceleration processes, which allow the electrons to 
extend their lifetimes, 
c) the presence of bulk streaming motions in the radio-emitting plasma. The 
existence of bulk streaming motions in radio galaxies has been suggested by 
Liu et al. (1989), on the basis of similar electron ageing arguments. 
We have also considered the possibility
that the magnetic field, which enters in the computation of the age,
 is not the equipartition value. We have noted, however, that 
the equipartition magnetic fields in these sources 
are always close to the values which give the maximum lifetimes
(Van der Laan \& Perola 1969), therefore the use of significantly 
different values of the magnetic field would make the velocities still higher.

\subsection{Polarization properties} 

We have determined the linear polarization in all sources at 10.6~GHz, and at 
4.8~GHz, and 2.7~GHz (when available).
The degree of polarization generally increases along the tail, reaching 
values larger than 50\% at 10.6~GHz. Towards the head, the degrees of 
polarization is lower, owing to beam depolarization within the regions where 
the twin jets (and hence the magnetic fields tied to them) get strongly bent 
backwards. For the same reason, the measured fractional polarization can also 
be significantly lower at the longer wavelengths. Synchrotron radiation theory 
predicts that the intrinsic polarization percentage is given by
$$ \rm P_{\rm intr} = {{3\alpha + 3} \over {3\alpha + 5}}~~~~ (2)$$
where $\alpha$ is the observed emission spectral index. At the tail's end the 
spectral index typically increases to $\alpha$$\sim$2; then the intrinsic 
fractional polarization expected there is $\sim$80~\%. The observed
polarization percentage is related to the intrinsic theoretical one by the 
expression 
$$ \rm P_{\rm obs} = P_{\rm intr} {{H^2_{\rm o}} \over {H^2_{\rm o} + 
H^2_{\rm r}} }~~~~ (3)$$
where H$_{\rm o}$ and H$_{\rm r}$  are the ordered and random component of the 
magnetic field, respectively. The high linear polarization percentage seen 
at 10.6~GHz implies a high, though not complete, degree of ordering of the 
magnetic field. The increase of the polarization percentage with distance from 
the core is much stronger than that expected from the spectral index variation 
(Eq.~2). Therefore, the degree of ordering of the magnetic field is higher as 
the radio emitting plasma is getting 
older. In 3C\ts129, the degree of polarization at the 
tail's end is close to the theoretical value. We derive 
that 91\% of the magnetic field in this region 
is ordered on a scale of $\sim$20~kpc. At $\sim$5$^{\prime}$ 
from the core, the  ordered component of the magnetic field is 
$\sim$65\%. In the 
other sources, the uniform component of the magnetic field is $\sim$70--75\% 
at the tails' ends on a 20--30~kpc scale.
If the increase in the ordering of the magnetic field is due to reconnection
of the magnetic field (e.g. Soker \&  Sarazin 1990), then this would be
accompanied by energy release which could also be responsible for
reaccelerating the electrons.

In all sources, the magnetic field is aligned with the tails. This alignment 
has been interpreted as due to the anisotropic effect of ram pressure
(Pacholczyk \& Scott 1976) and is consistent with the increase of magnetic
field ordering along the tails. 

\section{Conclusions}

We have presented a multi-frequency study of 4 tailed radio galaxies. The 
results are summarized in the following.

Point-to-point spectra using 3 or 4 frequencies, in the range from 327~MHz to 
10.6~GHz, have been obtained along the tails, up to the outermost regions. 
We find that the spectrum in each radio galaxy progressively steepens with 
distance from the core, and at each location it is steeper at higher  
frequencies. Except for the N-S tail of 3C\ts465, the spectra 
are well fitted by models involving synchrotron 
energy losses, assuming both that the electron pitch angle is constant over 
time (KP model), and that there is continuous 
isotropization of the pitch-angle  
distribution (JP model). The spectral fits do not allow us to argue in favour
of either the KP or JP model.
The critical frequency obtained by spectral fits 
decreases with increasing distance from the core. Assuming that the magnetic 
field is at the equipartition value, we show that the lifetimes of radiating 
electrons after synchrotron and Inverse Compton losses are generally between 
10$^7$ and 10$^8$~yr, with the exception of IC\ts310 where the 
electrons in the 
outer tail are older than 10$^8$~yr. Electron drift velocities are generally
larger than 2000~km/s, and show significant variations 
along the tails. This implies that 
reacceleration processes or bulk motions should be present. 

The polarization data at 10.6~GHz indicate that in all the sources
the magnetic fields are 
oriented along the tails. The polarization percentage increases along the 
tails, reaching values of $\simeq$50\%. In 3C\ts129 the polarization 
percentage in the outermost regions is $\sim$75\%, close to the theoretical 
synchrotron value. The degree of ordering of the magnetic field increases 
along the tails, viz. up to $\sim$75\% in IC\ts310, NGC\ts1265, and 3C\ts465. 
In 3C\ts129 the ordered component of the magnetic field exceeds 90\% at the 
tail's end. 

\begin{acknowledgements}
We thank Matteo Murgia for his help in the application of the code
for the spectral fits. We are grateful to the referee,  Chris O'Dea,
for helpful suggestions. Part of this work was supported by the Deutsche
Forschungsgemeinschaft, grant KL533/4-2 and by European Commission, TMR
Programme, Research Network Contract ERBFMRXCT97-0034 ``CERES''.
\end{acknowledgements}

\begin{figure*}
\includegraphics{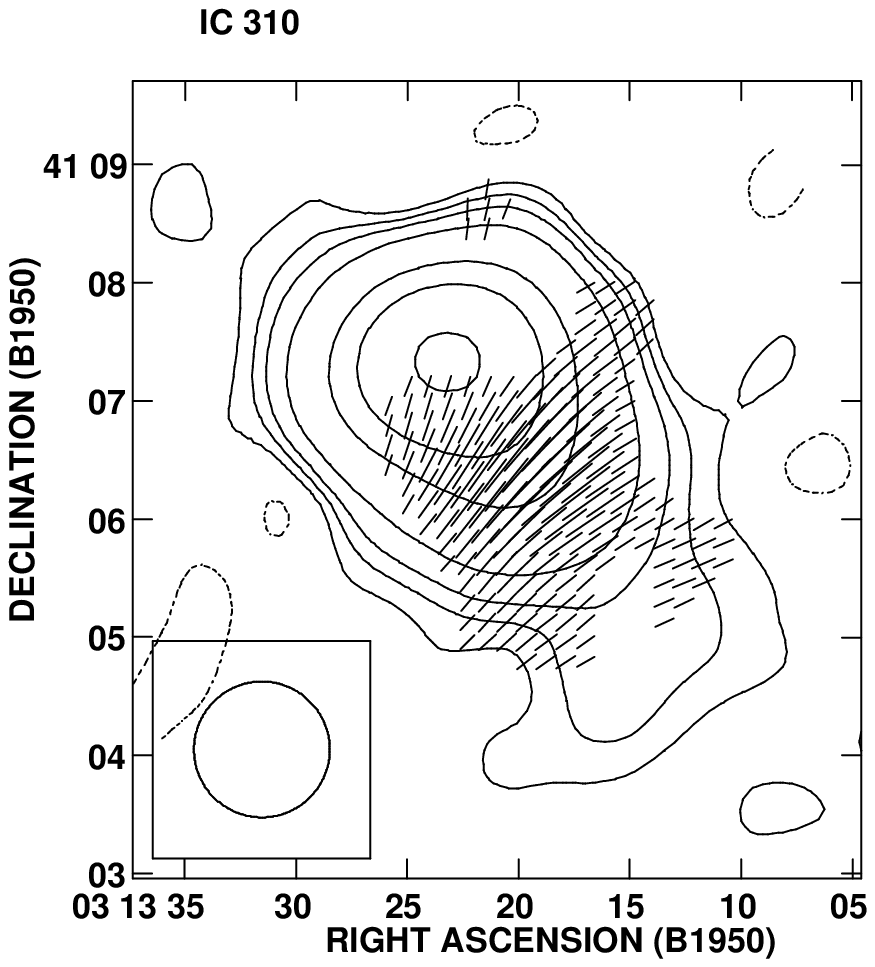}
\vspace{10 cm}
%\picplace{1.0cm}    %   spazio che lascia per la figura 
\caption{ Contour map of IC 310 at 10.6 GHz. Superimposed lines respresent the
orientation of the polarization vector and are proportional in length to the
polarized intensity (1$^{\prime \prime}$ = 0.1 mJy/beam).
Contour levels are: -1, 1.5, 3, 5, 10, 30, 50, 90 mJy/beam.} 
\end{figure*}%  dentro graffa per figura larga tutta la pagina 

\begin{figure*}%  * dentro graffa per figura larga tutta la pagina
\includegraphics{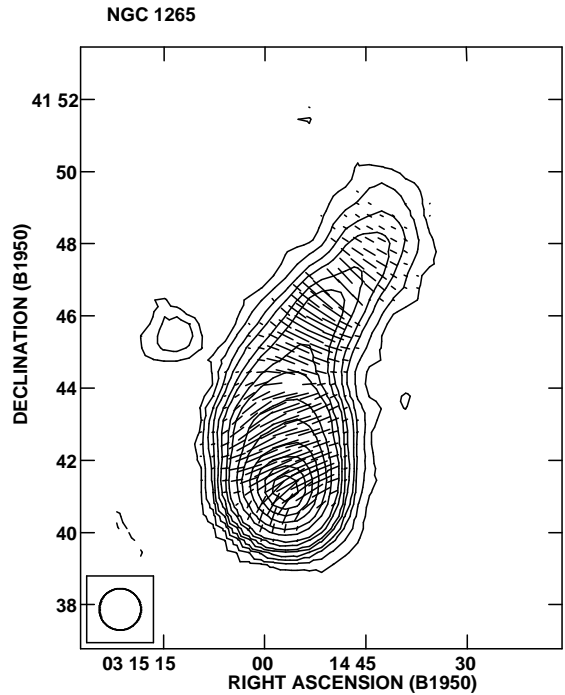}
\vspace{10 cm}
%\picplace{1.0cm}    %   spazio che lascia per la figura 
\caption{ Contour map of NGC 1265  at 10.6 GHz. 
The peak to S-E of the source in the 2.7 GHz image is the unrelated background
source 3C\ts83.1A. 
The polarization vector and are proportional in length to the
polarized intensity, with 1$^{\prime \prime}$ = 0.25 mJy/beam.
Contours levels are as follows: 2, 4, 7, 10, 15, 20, 30, 40, 70, 100,
150, 200, 250, 300 mJy/beam.}
\end{figure*}%  dentro graffa per figura larga tutta la pagina 

\begin{figure*}
\includegraphics{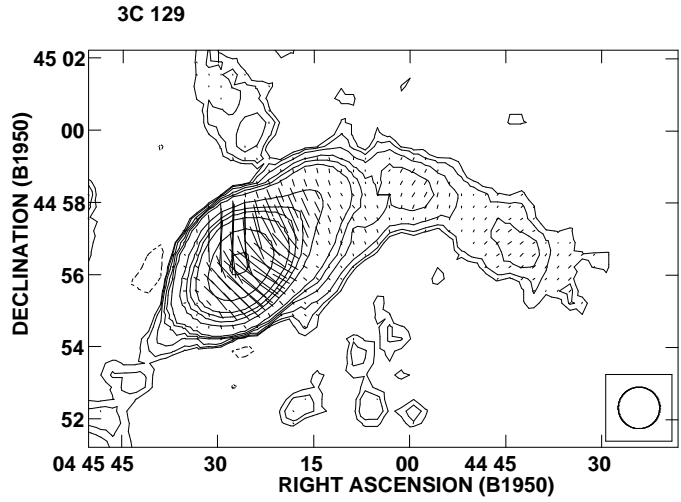}
\vspace{10 cm}
%\picplace{1.0cm}    %   spazio che lascia per la figura 
\caption{ Contour map of 3C 129 at 10.6
GHz. Superimposed lines represent the orientation of the
polarization vector and are proportional in length to the polarized intensity,
with 1$^{\prime \prime}$ = 0.3 mJy/beam. Contour levels are: 
-1.5, 1.5, 2, 3, 5, 7, 10, 30, 50, 70, 100, 200, 400 mJy/beam.}
\end{figure*}%  dentro graffa per figura larga tutta la pagina 

\begin{figure*}
\includegraphics{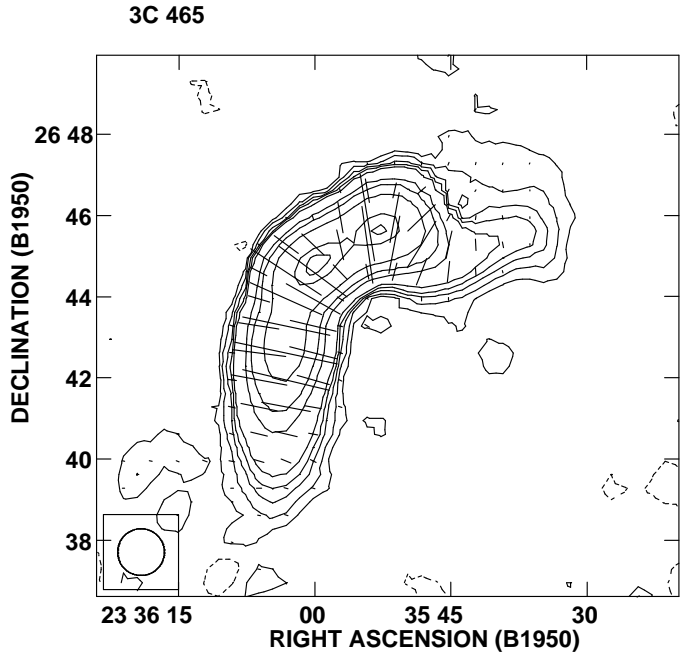}
\vspace{10 cm}
%\picplace{1.0cm}    %   spazio che lascia per la figura 
\caption{ Contour map of 3C 465 at 10.6 GHz.
The polarization vectors 
are proportional in length to the polarized intensity, with 
1$^{\prime \prime}$ = 0.3 mJy/beam. Contour levels are as follows: 
-1.5, 1.5, 3, 5, 7, 10, 30, 50, 100, 250, 300 mJy/beam.} 
\end{figure*}%  dentro graffa per figura larga tutta la pagina 

\begin{figure*}
\includegraphics{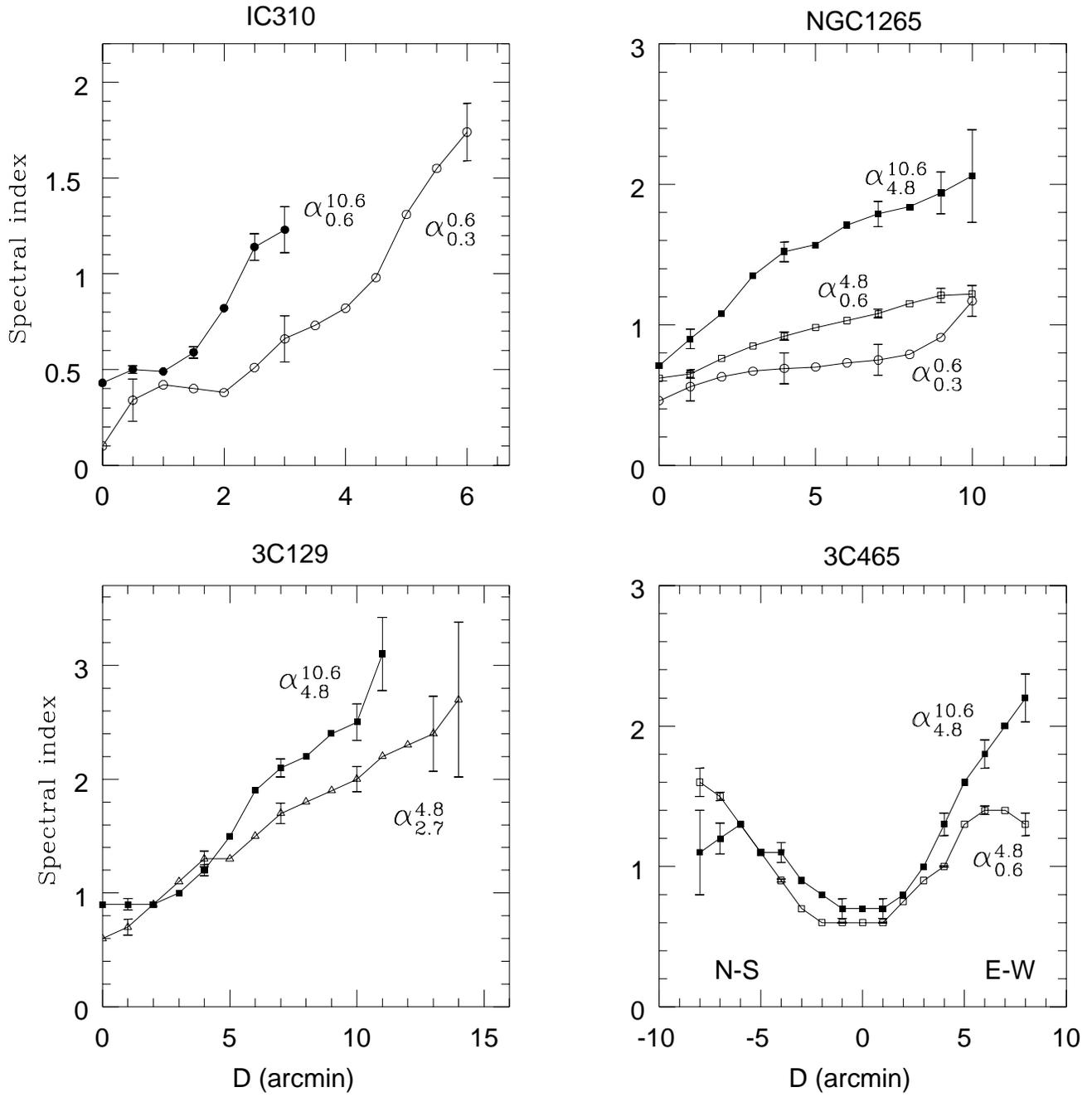}
\vspace{22 cm}
%\picplace{1.0cm}    %   spazio che lascia per la figura 
\caption{Trend of the spectral index of the sources under 
study along the ridge of maximum brightness.} 
\end{figure*}%  dentro graffa per figura larga tutta la pagina 

\begin{figure*}
\includegraphics{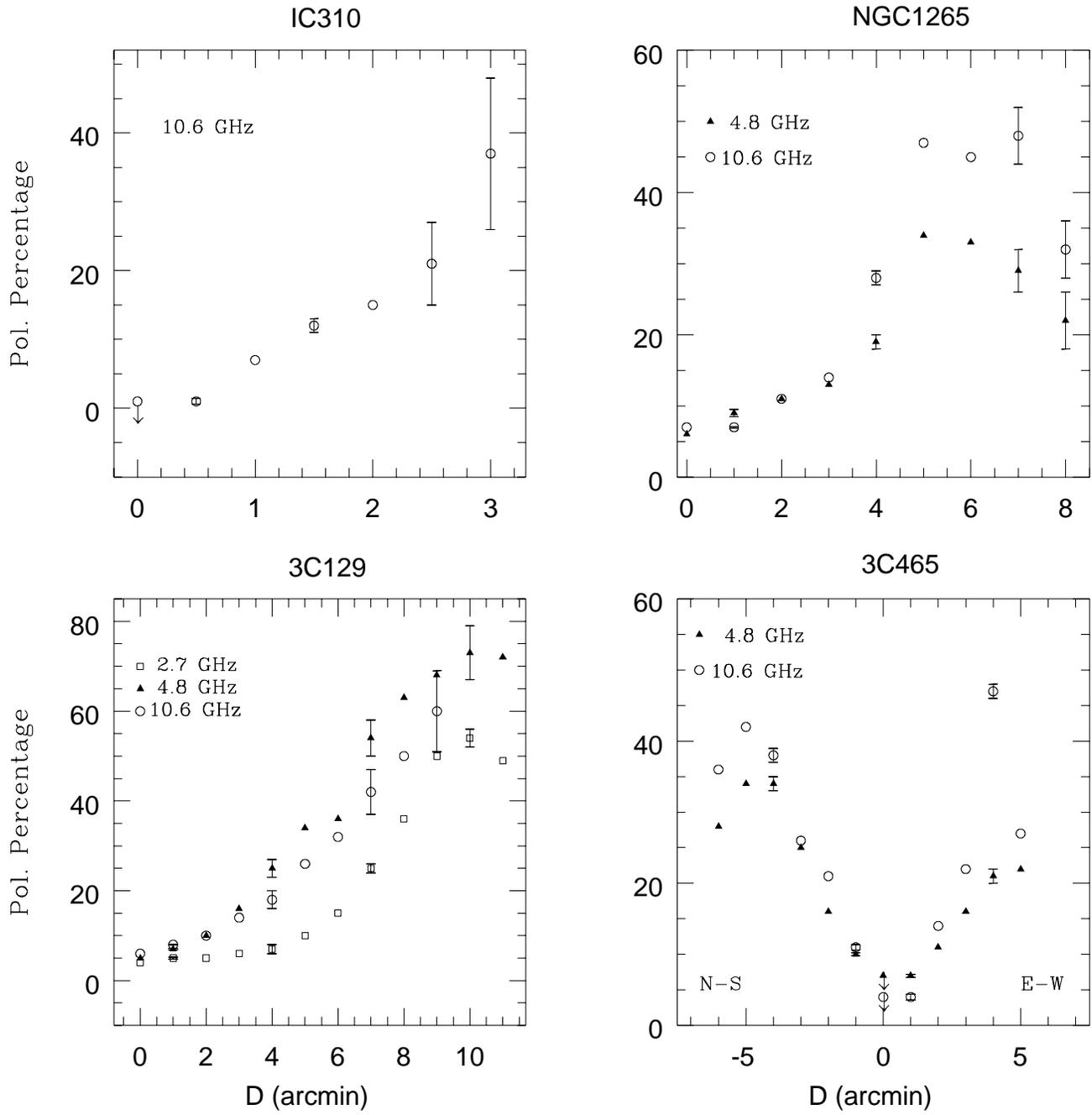}
\vspace{22 cm}
%\picplace{1.0cm}    %   spazio che lascia per la figura 
\caption{ Polarization percentage of the sources under study along the ridge of
maximum brightness. }
\end{figure*}%  dentro graffa per figura larga tutta la pagina 

\begin{figure*}
\includegraphics{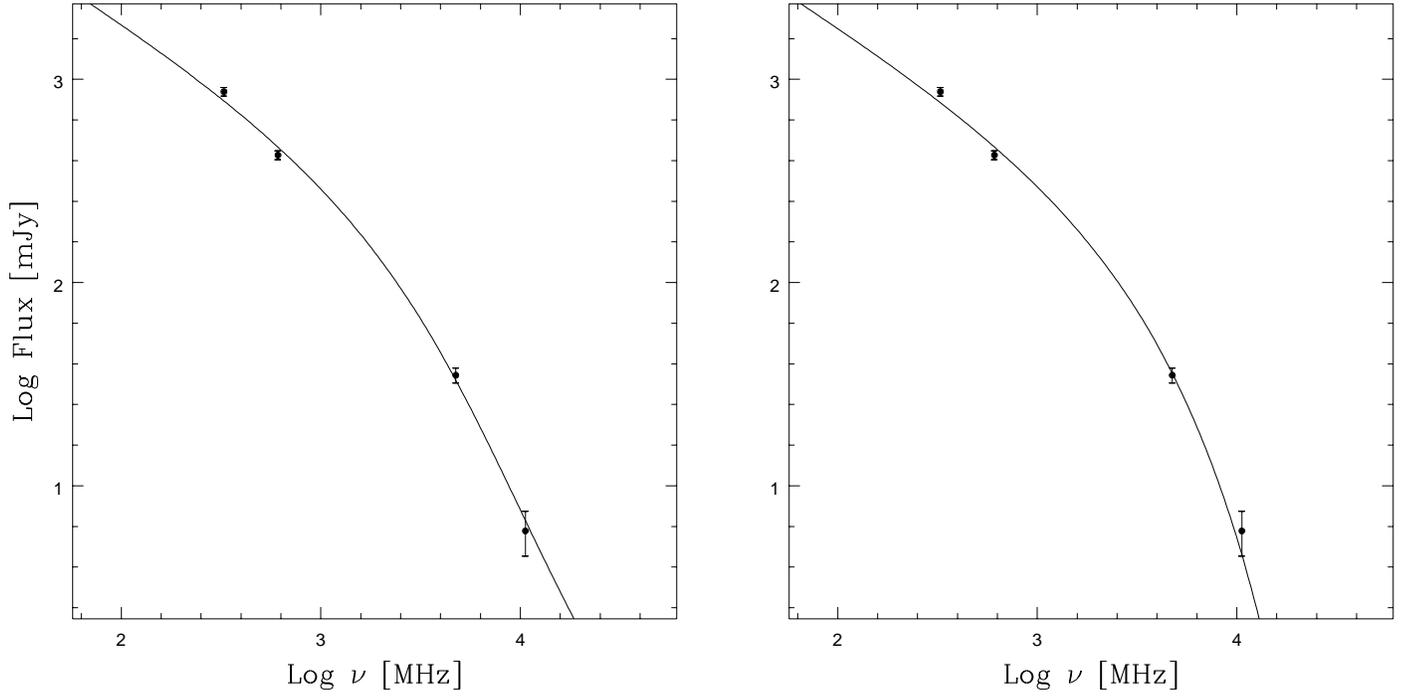}
\vspace{10 cm}
\caption{ Examples of the fit to the spectrum of NGC 1265
at 10$^{\prime}$ from the core. 
The continuous line represents the best fit model, obtained
 with $\gamma$ = 2.2. 
The left and right panels refer to the KP and JP  model, respectively. }

\end{figure*}%  dentro graffa per figura larga tutta la pagina 

\begin{figure*}
%\special{psfile=fig8a.ps hoffset=40 voffset=-490 hscale=65 vscale=65 }
%\special{psfile=fig8b.ps hoffset=40 voffset=-670 hscale=65 vscale=65 }
\includegraphics{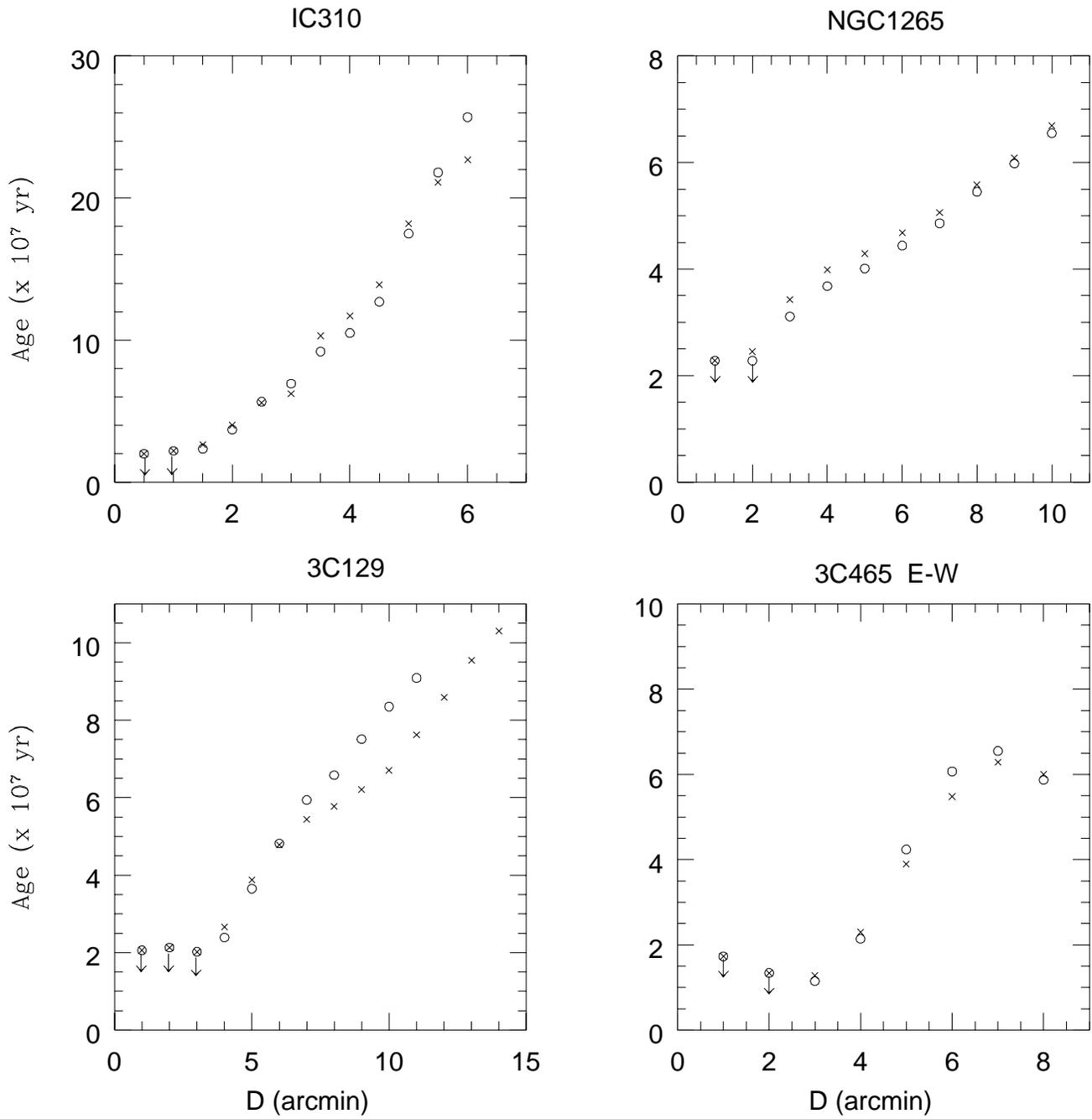}
\vspace{22 cm}
%\picplace{1.0cm}    %   spazio che lascia per la figura 
\caption{ Plots of the radiating electron lifetimes  at different
distance from the core. The open circles and crosses refer to the
Kardashev-Pacholczyk and Jaffe-Perola model, respectively.}
\end{figure*}%  dentro graffa per figura larga tutta la pagina 

\end{document}